\documentclass[sigconf]{acmart}

\copyrightyear{2024}
\acmYear{2024}
\setcopyright{acmlicensed}\acmConference[CIKM '24]{Proceedings of the 33rd ACM International Conference on Information and Knowledge Management}{October 21--25, 2024}{Boise, ID, USA}
\acmBooktitle{Proceedings of the 33rd ACM International Conference on Information and Knowledge Management (CIKM '24), October 21--25, 2024, Boise, ID, USA}
\acmDOI{10.1145/3627673.3679704}
\acmISBN{979-8-4007-0436-9/24/10}
\usepackage{algorithmic}
\usepackage{algorithm}
\usepackage{multirow}
\usepackage{arydshln}
\usepackage{bm}
\usepackage{subcaption}
\usepackage{graphicx}
\usepackage{enumitem}
\usepackage{balance}

\begin{document}

\title{Exploiting Pre-trained Models for Drug Target Affinity Prediction with Nearest Neighbors}

\author{Qizhi Pei}
\authornote{Both authors contributed equally to this research.}
\affiliation{%
  \institution{Gaoling School of AI (GSAI),\\ Renmin University of China}
  \city{Beijing}
  \country{China}
}
\email{qizhipei@ruc.edu.cn}
\orcid{0000-0002-7242-422X}

\author{Lijun Wu}
\authornotemark[1]
\authornote{Corresponding authors: Rui Yan (ruiyan@ruc.edu.cn) and Lijun Wu (lijun\_wu@outlook.com).}
\affiliation{%
  \institution{Microsoft Research}
  \city{Beijing}
  \country{China}}
\email{lijun_wu@outlook.com}
\orcid{0000-0002-3530-590X}

\author{Zhenyu He}
\affiliation{%
  \institution{School of Intelligence Science and Technology, Peking University}
  \city{Beijing}
  \country{China}}
\email{hezhenyu@stu.pku.edu.cn}
\orcid{0009-0005-7001-0591}

\author{Jinhua Zhu}
\affiliation{%
  \institution{University of Science and Technology of China}
  \city{Hefei}
  \state{Anhui}
  \country{China}}
\email{teslazhu@mail.ustc.edu.cn}
\orcid{0000-0003-2157-9077}

\author{Yingce Xia}
\affiliation{%
  \institution{Microsoft Research}
  \city{Beijing}
  \country{China}}
\email{yingce.xia@microsoft.com}
\orcid{0000-0001-9823-9033}

\author{Shufang Xie}
\affiliation{%
  \institution{Gaoling School of AI (GSAI),\\ Renmin University of China}
  \city{Beijing}
  \country{China}
}
\email{shufangxie@ruc.edu.cn}
\orcid{0000-0002-7126-0139}

\author{Rui Yan}
\authornotemark[2]
\affiliation{%
  \institution{Gaoling School of AI (GSAI), \\ Renmin University of China \& \\ Engineering Research Center of Next-Generation Intelligent Search and Recommendation, Ministry of Education}
  \city{Beijing}
  \country{China}
}
\email{ruiyan@ruc.edu.cn}
\orcid{0000-0002-3356-6823}

\renewcommand{\shortauthors}{Qizhi Pei et al.}

\begin{abstract}
Drug-Target binding Affinity (DTA) prediction is essential for drug discovery. 
Despite the application of deep learning methods to DTA prediction, the achieved accuracy remain suboptimal. In this work, inspired by the recent success of retrieval methods, we propose $k$NN-DTA, a non-parametric embedding-based retrieval method adopted on a pre-trained DTA prediction model, which can extend the power of the DTA model with no or negligible cost. Different from existing methods, we introduce two neighbor aggregation ways from both embedding space and label space that are integrated into a unified framework. Specifically, we propose a \emph{label aggregation} with \emph{pair-wise retrieval} and a \emph{representation aggregation} with \emph{point-wise retrieval} of the nearest neighbors. This method executes in the inference phase and can efficiently boost the DTA prediction performance with no training cost. In addition, we propose an extension, Ada-$k$NN-DTA, an instance-wise and adaptive aggregation with lightweight learning. Results on four benchmark datasets show that $k$NN-DTA brings significant improvements, outperforming previous state-of-the-art (SOTA) results, e.g, on BindingDB IC$_{50}$ and $K_i$ testbeds, $k$NN-DTA obtains new records of RMSE $\bf{0.684}$ and $\bf{0.750}$. The extended Ada-$k$NN-DTA further improves the performance to be $\bf{0.675}$ and $\bf{0.735}$ RMSE. These results strongly prove the effectiveness of our method. Results in other settings and comprehensive studies/analyses also show the great potential of our $k$NN-DTA approach.
\end{abstract}





\maketitle

\section{Introduction}
Drug discovery has been more and more important, which is a long and expensive process that typically takes tens of years and billions of dollars. Therefore, Computer-Aided Drug Discovery (CADD) plays an important role to help accelerate the journey, especially in the early stage. Among various CADD applications, Drug-Target binding Affinity (DTA) prediction is an essential one. DTA measures the interaction strength between a drug and a target, and the accurate prediction can greatly benefit Virtual Screening (VS)~\citep{inglese2007high} and expedite drug repurposing~\citep{pushpakom2019drug}, e.g., finding potential drugs for COVID-19~\citep{zhou2020artificial}. Along the way, various computational methods have been proposed for DTA prediction~\citep{gilson2007calculation,trott2010autodock,salsbury2010molecular,pahikkala2015toward}.

Recently, Deep Learning (DL) methods have been widely applied for DTA prediction with the increased available affinity data, and a huge process has been made~\citep{ozturk2018deepdta,nguyen2021graphdta,huang2020deeppurpose}. Though DL-based methods are popular, DTA is still an unsolved problem with unsatisfied accuracy~\cite{d2020machine}. Besides, the training cost of DL-based DTA methods is still high (e.g., multi-GPUs with tens of hours or days) and there are many different deep models already trained for DTA prediction. Therefore, we are thinking of the following question, \emph{how can we further exploit the potential of these existing DL-based DTA models with no or little effort?}

Luckily, non-parametric methods (e.g., $k$-nearest neighbors) have shown success in various tasks recently, such as language modeling~\citep{khandelwal2019generalization}, machine translation~\citep{gu2018search,khandelwal2020nearest}, question answering~\cite{guu2020realm}, and retrosynthesis~\cite{xie2023retrosynthesis}. These methods have demonstrated their effectiveness by making the neural models expressive, adaptable, and interpretable. Therefore, in this paper, we propose $k$NN-DTA as a solution to answer the above question, which is an embedding-based non-parametric approach that utilizes nearest neighbors for DL-based DTA prediction. That is, we utilize the drug and target representations extracted from the trained models to retrieve the nearest samples in a datastore (e.g., the original training data).  Compared with traditional chemical similarity retrieval, our embedding-based retrieval has much higher efficiency and also quality guarantee (see Section~\ref{sec:chemical_similarity}).  Different from common approaches, $k$NN-DTA introduces two aggregation ways for the retrieved neighbors from both label and embedding spaces. Specifically, a \emph{label aggregation} is performed on the nearest neighbors of drug-target pairs with a \emph{pair-wise embedding retrieval}. Besides, a \emph{representation aggregation} is also conducted on the nearest drug or target representations with a \emph{point-wise embedding retrieval}. The integrated labels and the model prediction are then combined as the final affinity score. Note that $k$NN-DTA only needs to execute in the inference phase for a pre-trained DTA model, hence it boosts affinity predictions without any extra training in an efficient and effective way. The so-called ``pre-trained DTA'' is the model directly trained on the DTA task without further fine-tuning, which differs from the typical pre-training and fine-tuning.

We further introduce an extension of $k$NN-DTA, Ada-$k$NN-DTA, with lightweight training cost. In Ada-$k$NN-DTA, a plug-and-play learning module is designed, where the neighbor distances are taken as input to obtain adaptive and instance-wise weights for aggregation. The intuitive motivation behind this is that, since each data sample has different neighbors w.r.t the embedding/label distance closeness, adaptive aggregation can potentially boost more precise prediction from these neighbors. Besides, the light training module can automatically learn how to aggregate the neighbors so to avoid manual hyperparameter tuning costs.

We conduct extensive experiments on four benchmarks for evaluation, including BindingDB IC$_{50}$ and $K_i$, DAVIS and KIBA datasets. On all datasets, significant performance improvement is achieved by $k$NN-DTA against pre-trained models, and new state-of-the-art (SOTA) results are obtained. For example, on BindingDB IC$_{50}$ and $K_i$ testbeds, $k$NN-DTA reaches the new best RMSE, $\bf{0.684}$ and $\bf{0.750}$. With Ada-$k$NN-DTA, the prediction error is further reduced by about $0.01$ RMSE. We then test on four generalization testsets through zero-shot transfer learning, in which $k$NN-DTA also demonstrates its potential generalization ability. At last, we also deeply show the effectiveness of $k$NN-DTA and Ada-$k$NN-DTA by conducting comprehensive studies. 

The contributions of this work are as follows. (1) We propose $k$NN-DTA, a novel non-parametric embedding-based retrieval method to exploit the great potential of existing DL-based DTA models with minor or no additional training, which includes two proposed aggregation ways from both embedding space and label space with different retrieval methods. (2) We further introduce an extension of a lightweight Ada-$k$NN-DTA framework to learn adaptive aggregations with little cost. (3) We conduct extensive experiments and comprehensive studies to demonstrate the effectiveness and high efficiency of our approaches, and new SOTA results are achieved among various testbeds. (4) Lastly, since affinity prediction is highly crucial for virtual screening so as to efficiently select potential drugs, our paper delivers the message to chemists/data scientists that using the embedding retrieval method upon deep models is a good way to do DTA prediction. We hope our approach can benefit/inspire more people (especially in AI4Science) to think along this way and do more advanced innovations.

\section{Related Work}

\noindent{\textbf{Drug-Target binding Affinity (DTA) Prediction}} aims to estimate the strength of drug-target interaction.
The experimental assay~\citep{inglese2007high} is the most reliable method, but it is labor-intense with high cost.
Hence, computational methods have been applied, which can be divided into structure-based and structure-free methods.
For structure-based ways, molecular docking~\citep{trott2010autodock,verdonk2003improved} and molecular dynamics simulations~\citep{salsbury2010molecular} are typical ones. 
For structure-free methods, machine learning ways include Random Forest~\citep{shar2016pred}, kernel-based works~\citep{cichonska2017computational}, and gradient boosting machines~\citep{he2017simboost}.
Thanks to the increased available affinity data, deep learning models~\citep{ozturk2018deepdta,nguyen2021graphdta,ozturk2019widedta,huang2020deeppurpose,nguyen2022gefa} are now dominating the DTA prediction, which takes different neural networks (e.g., GNNs) for representation learning. 

\noindent{\textbf{Similarity-based Virtual Screening}} is commonly adopted in classical binding prediction, which usually generates drug and target similarity matrices~\citep{thafar2022affinity2vec,abbasi2020deepcda,ding2014similarity,shim2021prediction,shi2018inferring,ru2022nerltr,islam2021dti}.
These similarity matrices serve as features to be integrated into different methods, such as kernel regression~\citep{yamanishi2008prediction}, matrix factorization~\citep{bolgar2016bayesian}, gradient boosting machine~\citep{tanoori2021drug,thafar2021dti2vec}, neural network classifiers~\citep{shim2021prediction,an2021heterogeneous}, and so on. 
Several works also utilize the drug/target similarity to integrate the affinity labels~\cite{van2013predicting,liu2022drug}. 
SVM-KNN~\citep{zhang2006svm} is a work that combines the $k$NN with an SVM classifier for prediction, but it differs a lot from ours on motivation and process. 

\noindent{\textbf{Nearest Neighbor Learning and Memory Networks.}
Recently, $k$NN retrieval is popular in Natural Language Processing (NLP)~\citep{kaiser2017learning,gu2018search,borgeaud2021improving,zheng2021adaptive,lewis2020retrieval}.
\cite{khandelwal2019generalization} is among the first work that successfully combines the language model with $k$NN retrieval method. 
Later, \cite{khandelwal2020nearest} shares a similar idea to apply $k$NN retrieval to machine translation. 
After that, the $k$NN-based methods are widely spread in different areas, such as question answering~\citep{lewis2020retrieval}, pre-training~\citep{guu2020realm,guu2020retrieval}, and dialogue conversation~\citep{fan2021augmenting}.
Our $k$NN-DTA is inspired by them but differs from the aggregation methods and the regression prediction scenarios. Another related field is Memory Networks~\citep{weston2015memory,sukhbaatar2015end,zhang2017dynamic} which utilizes an explicit memory module. However, memory networks are part of the model and must be trained and updated, which are mainly designed for LSTM~\citep{hochreiter1997long} to extend the memory cell.}

\section{Method}
In this section, we first define the DTA prediction task and necessary notations. Next, we introduce our $k$NN-DTA with proposed two aggregation and retrieval ways. Then, we present the extension Ada-$k$NN-DTA. Finally, we give some discussions.

\noindent{\textit{Preliminary}.}
Let $\mathcal{D} = \{(D, T, y)_i\}_{i=1}^N$ denotes a DTA dataset, where $(D, T, y)$ is a triplet sample and $N$ is the dataset size. Here $D/T$ is one drug/target from the dataset, and $y$ (a floating number) is the label measuring the binding affinity strength (e.g., IC$_{50}$, $K_i$, $K_d$) between the drug-target pair. The DTA prediction is then a regression task that aims to predict the affinity score between the drug-target pair. Mathematically, the goal is to learn a mapping function $\mathcal{F}: D \times T \to y$. 
A drug $D$ can be represented by different formats, such as simplified molecular-input line-entry system (SMILES)~\citep{weininger1988smiles}, or graph with nodes (atoms) and edges (bonds), or a 3D conformation where the coordinates of all atoms are available. Similarly, a target $T$ can be represented by amino acid sequences or a 3D conformation. 
In this work, we take the SMILES strings for drug and amino acid sequences for target. 
Due to the superior performance of Transformer~\citep{vaswani2017attention}, we use two Transformer encoders $\mathcal{M_D}$ and $\mathcal{M_T}$ to encode the drug $D$ and target $T$ and obtain $R_D$ and $R_T$. The $R_D$ and $R_T$ are fed into a prediction module $\mathcal{P}$ to get the predicted affinity $\hat{y}$.

\begin{figure*}[tbp]
    \centering
    % \vspace{-0.2cm}
    \advance\leftskip-0.5cm
    \includegraphics[scale=0.58]{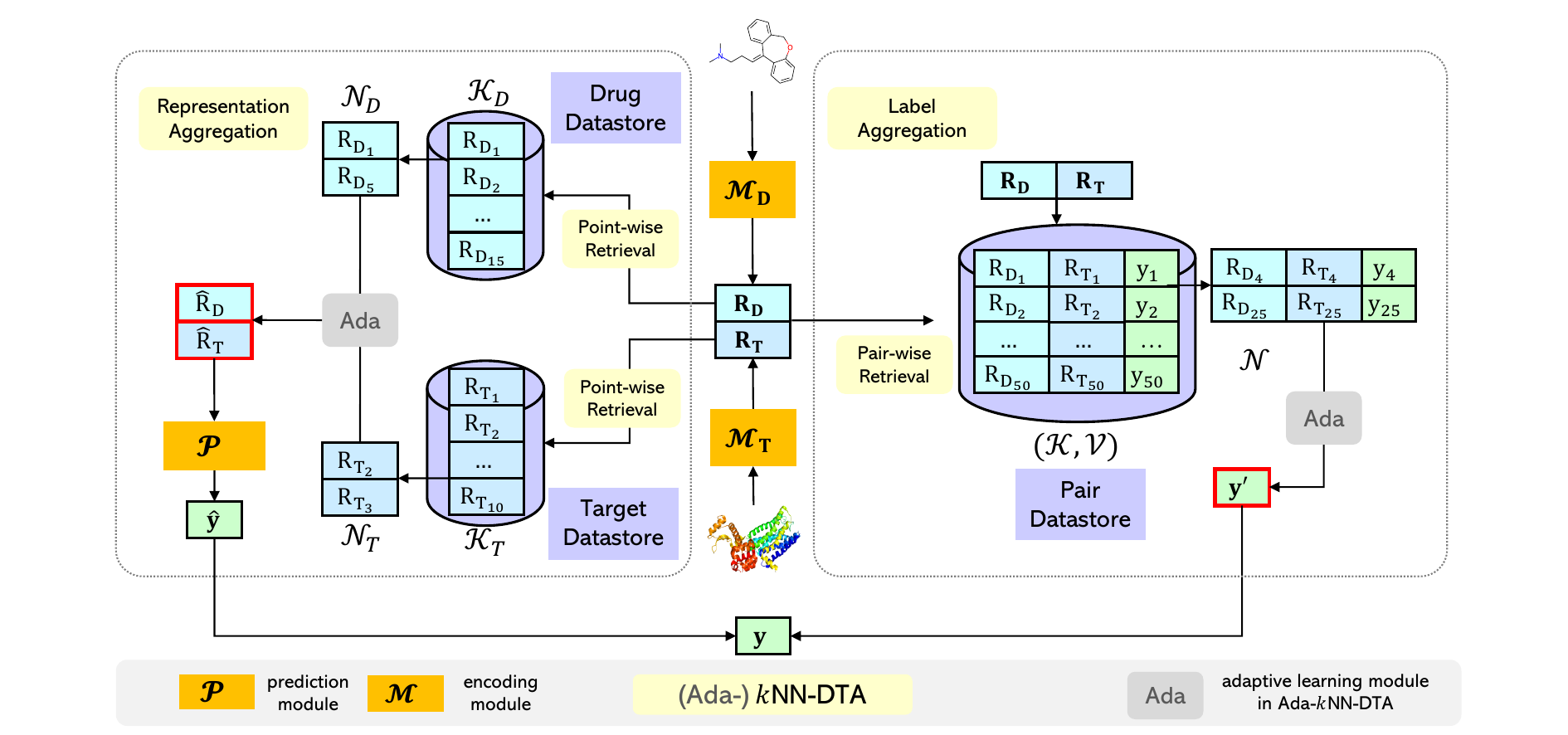}
    \vspace{-0.7cm}
    \caption{The overall framework of our $k$NN-DTA and Ada-$k$NN-DTA. We use two Transformer encoders $\mathcal{M}_D$ and $\mathcal{M}_T$ to encode drug $D$ and target $T$. The representations $R_D$ and $R_T$ are separately used for representation aggregation with point-wise retrieval. Meanwhile, the concatenation of $R_{D}$ and $R_{T}$ are then used for label aggregation with pair-wise retrieval. The dashed grey `Ada' parts are the lightweight learning modules in Ada-$k$NN-DTA. `$\mathcal{P}$' stands for the prediction module, $\mathcal{(K, V)}$, $\mathcal{K}_D$, $\mathcal{K}_T$ are the datastores, and $\mathcal{N}$, $\mathcal{N}_D$, $\mathcal{N}_T$ are retrieved nearest neighbors. The aggregated representation and the affinity are in red outline.}
    \label{fig:pipeline}
    \vspace{-0.3cm}
\end{figure*}

\subsection{Retrieval-based $k$NN-DTA}
Our $k$NN-DTA incorporates two retrieval methods and aggreation ways, which are label aggregation with pair-wise retrieval and representation aggregation with point-wise retrieval.

\subsubsection{Label Aggregation with Pair-wise Retrieval}
Intuitively, similar drug-target pairs possibly have similar binding affinity scores. Hence, we propose \emph{label aggregation} with \emph{pair-wise retrieval}, which is to aggregate the ground-truth affinity scores from $k$ nearest neighbors retrieved by the embeddings of drug-target pair. 
Shortly speaking, we first build a key-value memory datastore that contains the encoded representations of all drug-target pairs and their corresponding labeled affinity values, which can be quickly done through a single forward pass of the pre-trained DTA model. Then, the $k$NN retrieval is performed when evaluating on test samples.

\paragraph{Datastore} 
The memory datastore is constructed offline with a set of key-value pairs $(k_i, v_i)$. Since the affinity score $y$ corresponds to a specific drug-target pair $(D_i, T_i)$ instead of one drug or target only, the key $k_i$ in our datastore is the concatenated representation of $R_{D_i}$ and $R_{T_i}$, that is $[R_{D_i}; R_{T_i}]$, and the value $v_i$ is the ground-truth affinity score $y_i$. This is the why we call {\em pair-wise} retrieval.
The datastore $(\mathcal{K}, \mathcal{V})$ is created by the key-value pairs for all the samples in dataset $\mathcal{D}$,
\begin{equation}
    (\mathcal{K}, \mathcal{V}) = \{([R_{D_i}; R_{T_i}], y_i) | ((D_i, T_i), y_i) \in \mathcal{D})\}.
\end{equation}
Noting that we only need a single forward pass of the pre-trained DTA model to obtain $(\mathcal{K}, \mathcal{V})$, which can be quickly done.

\paragraph{Pair-wise Retrieval, Label Aggregation, Affinity Prediction}
Given a test sample $(D_t, T_t)$, we first encode the data through encoder $\mathcal{M}_D$ and $\mathcal{M}_T$ to obtain representations $R_{D_t}$ and $R_{T_t}$. Then the concatenated $[R_{D_t}; R_{T_t}]$ is used as query to retrieve the $k$ nearest neighbors $\mathcal{N}=\{(k_i, v_i)\}=\{([R_{D_i}; R_{T_i}], y_i)\}$ from the datastore. The retrieval depends on specific similarity measurement $s(\cdot, \cdot)$ between query and the datastore, such as $L_2$ distance. 
With the retrieved nearest neighbor set, we then do \emph{label aggregation} among the labeled affinity from the neighbors in an attentive way. That is, a softmax is performed on the similarities $s([R_{D_t}; R_{T_t}], k_i)$ and we aggregate the retrieved affinity values $y_i$ by the attention weights to be $y'_t$. Mathematically, 
\begin{equation}
\label{eqn:label_integration}
    y'_t = \sum_{(k_i, v_i) \in \mathcal{N}} \alpha_i * y_i, \quad
    \alpha_i = \frac{\exp{(s([R_{D_t}; R_{T_t}], k_i)/\tau)}}{\sum_{(k_j, v_j) \in \mathcal{N}} \exp{(s([R_{D_t}; R_{T_t}], k_j)/\tau)}},
\end{equation}
where $\tau$ is the temperature, and $y_i$ equals to $v_i$ in above equations. The integrated affinity score $y_t$ is supposed to produce a good prediction with the help of retrieved neighbors. Since the pre-trained model $\mathcal{P}$ can also produce a prediction $\hat{y}_t$, we can further aggregate the aggregated affinity score $y'_t$ and the model prediction $\hat{y}_t$ as the final one, $y_t$ = $\lambda * \hat{y}_t + (1-\lambda) * y'_t$, where $\lambda$ is the coefficient.

\subsubsection{Representation Aggregation with Point-wise Retrieval}
Apart from above label aggregation that directly affects the predicted affinity scores through the label space, we also introduce another \emph{representation aggregation} with \emph{point-wise retrieval} to leverage the nearest neighbors from the embedding space. This is related to the similarity-based VS methods. Different from the above pair-wise retrieval, here we use separate \emph{point-wise} retrieval for $k_D$ nearest drug representations and $k_T$ nearest target representations. Generally speaking, we build separate datastores for drugs and targets, with only the key (drug and target representations) saved in the datastore since the values we need is the same as the keys (also the drug/target representations). Then $k$NN retrieval is performed on test drug and target to aggregate representations.

\paragraph{Datastore, Point-wise Retrieval, Representation Aggregation, Affinity Prediction}
We build a datastore $\mathcal{K}_D$ for drugs and a $\mathcal{K}_T$ for targets. 
Instead of the key-value pairs, these datastores only save keys $k_{D_i}$ and $k_{T_i}$. That is, the encoded drug/target representation $R_{D_i}/R_{T_i}$ is stored in $\mathcal{K}_D/\mathcal{K}_T$. 
Noting that the $R_{D_i}$ and $R_{T_i}$ are the same as that in above pair-wise retrieval method.
Thus $\mathcal{K}_D = \{R_{D_i} | D_i \in \mathcal{D}\}, \mathcal{K}_T = \{R_{T_i} | T_i \in \mathcal{D}\}$,
where $D_i$ and $T_i$ are the unique drugs and targets. 

At test time, given the test sample$(D_t, T_t)$, we use $R_{D_t}$/$R_{T_t}$ as query to retrieve nearest representations from $\mathcal{K}_D$/$\mathcal{K}_T$ with similarity metric $s(\cdot, \cdot)$. 
The retrieved sets are $\mathcal{N}_D = \{R_{D_i}\}$ and $\mathcal{N}_T = \{R_{T_i}\}$. The $k$NN retrieval is also based on similarity metric $s(\cdot, \cdot)$ between query representation and the ones in datastore. 
With the retrieved sets $\mathcal{N}_D$ and $\mathcal{N}_T$, attentive representation aggregation is conducted. Same as the label aggregation, the representation aggregation is $R'_{D_t}  = \sum_{R_{D_i} \in \mathcal{N}_D} \alpha^D_i * R_{D_i}, \alpha^D_i = \frac{\exp{(s(R_{D_t}, R_{D_i})/\tau_D)}}{\sum_{R_{D_j} \in \mathcal{N}_D} \exp{(s(R_{D_t}, R_{D_j})/\tau_D)}}$, and $R'_{T_t}$ is calculated in a same way.
With $R'_{D_t}/R'_{T_t}$, we further aggregate them with query $R_{D_t}/R_{T_t}$ to obtain the final drug and target representation, $\hat{R}_{D_t} = \lambda_D * R_{D_t} + (1-\lambda_D) * R'_{D_t}$ and $\hat{R}_{T_t} = \lambda_T * R_{T_t} + (1-\lambda_T) * R'_{T_t}$, which are then inputted to the model $\mathcal{P}$ for affinity prediction $\hat{y}_t = \mathcal{P}(\hat{R}_{D_t}, \hat{R}_{T_t})$.

\subsubsection{Unified Framework}
Each of the above aggregation methods can be used to enhance DTA prediction. In order to make the best use of above two ways, we systematically combine them in a unified framework, which is shown in Figure~\ref{fig:pipeline}. 
Given the test sample $(D_t, T_t)$, the whole test process is as follows. 
(1) Use encoders $\mathcal{M}_D$ and $\mathcal{M}_T$ to obtain the representations $R_{D_t}$ and $R_{T_t}$;
(2) Concatenate $R_{D_t}$ and $R_{T_t}$ and use it as a query to retrieve the nearest samples from $\mathcal{(K, V)}$. The label aggregation is performed to the retrieved neighbors affinity values to obtain $y'_t$;
(3) Use $R_{D_t}$/$R_{T_t}$ as query to separately retrieve the nearest drug/target representations from $\mathcal{K}_D$/$\mathcal{K}_T$, and aggregate retrieved representations and the query representations to obtain $\hat{R}_{D_t}$/$\hat{R}_{T_t}$, then get model prediction $\hat{y}_t=\mathcal{P}(\hat{R}_{D_t},\hat{R}_{T_t})$;
(4) The $y'_t$ are then combined with the predicted $\hat{y}_t$ to produce the final affinity prediction $y_t = \lambda * \hat{y}_t + (1-\lambda) * y'_t$.

\subsection{Extension: Adaptive Retrieval-based Ada-$k$NN-DTA}
The above $k$NN-DTA only requires retrieving nereast neighbors in the inference phase, and the calculation of the aggregation is parameter-free and training-free. Though efficient, the coefficients for aggregation, e.g., $\lambda/\lambda_{D}$, are manually designed hyper-parameters in current $k$NN-DTA and shared for all the test data, without considering the aggregation quality for each specific sample. Hence, to further exploit the power of $k$NN-DTA and reduce the manually tuning cost of these hyperparameters, we propose an adaptive learning extension Ada-$k$NN-DTA. 

In Ada-$k$NN-DTA, some lightweight modules are introduced to meta-learn the aggregation weights, e.g.,  $\alpha/\alpha_D$, and the coefficients, e.g., $\lambda/\lambda_{D}$. 
Concretely, the embedding distances between the query and neighbors $s(\cdot, \cdot)$ are fed into a light meta-network to learn the weights/coefficients
and then perform the aggregation. Take the label aggregation as an example, these $k$ distances are put as a vector (denoted as $S = [s_1, ..., s_k]$) and then fed into a FFN with softmax to output the aggregation weights and coefficient,
\begin{align}
    \nonumber y_t = \alpha_0 * \hat{y}_t + \sum_{i=1}^k &\alpha_i * y_i, \quad
    \alpha_i = \text{softmax}(\text{FFN}(S))_i, \\
    \text{FFN}(S) &= \max(0, SW_1 + b_1)W_2 + b_2, 
\end{align}
where $W$ and $b$ are the learnable parameters. Specially, the output dimension of FFN is $k+1$. After softmax over $k+1$ values, the first $\alpha_0$ is the coefficient $\lambda$.
In this way, we automatically learn the coefficient $\lambda$ and adaptive weights $\alpha$ for aggregation. Noting that the training is only conducted on the valid set and then the trained meta-network is directly applied on the test set.

\begin{table*}[t]
  % \vspace{-0.8cm}
  \centering
  \caption{Performance evaluation of different approaches on the BindingDB IC$_{50}$ and $K_i$ datasets. The $\downarrow$ and $\uparrow$ indicate the directions of better performances. Results are derived from three random runs.}
  \label{tab:bindingdb_result}
%   \resizebox{0.95\linewidth}{!}{
  % \scalebox{0.9}{
  \begin{tabular}{lccccc}
      \toprule
      \textbf{Dataset} & \multicolumn{2}{c}{\bf{IC$\bf{_{50}}$}} & \multicolumn{2}{c}{$\bf{K_i}$} & \\
      \midrule
      \textbf{Method} & \textbf{RMSE$\downarrow$} & \textbf{R$\uparrow$} & \textbf{RMSE$\downarrow$} & \textbf{R$\uparrow$} \\
      \midrule
      Random Forest~\citep{karimi2019deepaffinity} & 0.910 & 0.780  & 0.970 & 0.780 \\
      DeepAffinity~\citep{karimi2019deepaffinity} & 0.780 & 0.840 & 0.840 & 0.840\\
      DeepDTA~\citep{ozturk2018deepdta} & 0.782 & 0.848 & - & - \\
      MONN~\citep{li2020monn} & 0.764 & 0.858 & - & - \\
      BACPI~\citep{li2022bacpi} & 0.740 & 0.860 & 0.800 & 0.860 \\
      SSM-DTA~\citep{pei2023breaking} & 0.712 & 0.878 & 0.792 & 0.863 \\
      \midrule
      Pre-trained model & 0.717 (0.0066) & 0.880 (0.0037) & 0.785 (0.0016) & 0.876 (0.0008) \\
      \hdashline
      \quad + $k$NN-DTA & 0.684 (0.0021) & 0.889 (0.0012) & 0.750 (0.0016) & 0.882 (0.0004) \\
      \quad + Ada-$k$NN-DTA & 0.675 (0.0004) & 0.889 (0.0000) & 0.735 (0.0021) & 0.884 (0.0008)\\
      \bottomrule
  \end{tabular}
  % }
  \vspace{-0.3cm}
\end{table*}

\subsection{Discussion}
We put some clarification and discussion here. 
(1) For nearest neighbor retrieval, chemists/biologists usually utilize the data-specific chemical similarity, such as Tanimoto similarity based on fingerprint~\citep{fligner2002modification} for molecule and sequence identity for protein. Though domain specific, we compare them with our embedding similarity retrieval (in Section~\ref{sec:chemical_similarity}). Results show that embedding-based similarity has not only much higher efficiency but also outstanding performances. Hence, this is highly valuable to prove the superiority of the neural embedding retrieval.
(2) Our $k$NN-DTA builds three datastores, e.g., the drug-target pair datastore $\mathcal{(K, V)}$ and the drug/target datastore $\mathcal{K}_D/\mathcal{K}_T$. Actually, the representations stored in $\mathcal{K}_D/\mathcal{K}_T$ are the same as the ones in paired $\mathcal{(K, V)}$ with only duplicates removing. Thus, only one forward pass is required for constructing these datastores. 
(3) Different hyperparameters (e.g., $\lambda$, $k$) need to be searched in $k$NN-DTA. To reduce the cost, we first separately search for the label and representation aggregations, then we slightly search near these best configurations of them for the unified framework. 
(4) When aggregating the nearest labels or representations, we use the similarity-based softmax for combination. The simplest method is to use average pooling. In our experiments, we find our attentive way is better than the average one.
(5) Our current method uses embeddings of drugs and targets for retrieval and aggregation. Since drugs and targets are from two different domains, there remains much possibility for better integration, e.g., interaction-based attentive aggregation, this would be an interesting future point. 
(6) Finally, the basic assumption of our work depends on the similarity of the drug and target in the datastore (same as similarity-based VS), and we have somehow shown the reason behind the success in Section~\ref{sec:embed_visual}. 

\begin{table*}[t]
  \centering
  \caption{Performance evaluation of different approaches on the DAVIS and KIBA datasets. For DeepPurpose*~\citep{huang2020deeppurpose}, we take the best model setting from the original paper reports. Results are derived from three random runs.}
  \label{tab:davis_kiba_result}
%   \resizebox{0.95\linewidth}{!}{
  % \scalebox{0.9}{
  \begin{tabular}{lccccc}
      \toprule
      \textbf{Dataset} & \multicolumn{2}{c}{\textbf{DAVIS}} & \multicolumn{2}{c}{\textbf{KIBA}} & \\
      \midrule
      \textbf{Method} & \textbf{MSE$\downarrow$} & \textbf{CI$\uparrow$} & \textbf{MSE$\downarrow$} & \textbf{CI$\uparrow$} \\
      \midrule
      KronRLS~\citep{pahikkala2015toward} & 0.329 (0.019) & 0.847 (0.006) & 0.852 (0.014) & 0.688 (0.003) \\
      GraphDTA~\citep{nguyen2021graphdta} & 0.263 (0.015) & 0.864 (0.007) & 0.183 (0.003) & 0.862 (0.005) \\
      DeepDTA~\citep{ozturk2018deepdta} & 0.262 (0.022) & 0.870 (0.003)  & 0.196 (0.008) & 0.864 (0.002) \\
    %   \midrule
      DeepPurpose*~\citep{huang2020deeppurpose} & 0.242 (0.009) & 0.881 (0.005) & 0.178 (0.002) & 0.872 (0.001) \\
      DeepCDA~\citep{abbasi2020deepcda} & 0.248 (-) & 0.891 (0.003) & 0.176 (-) & 0.889 (0.002) \\
      Affinity2Vec~\citep{thafar2022affinity2vec} & 0.240 (-) & 0.887 (-) & 0.111 (-) & 0.923 (-) \\
      WGNN-DTA~\citep{jiang2022sequence} & 0.214 (-) & 0.892 (-) & 0.149 (-) & 0.892 (-) \\
      SSM-DTA~\citep{pei2023breaking} & 0.219 (0.001) & 0.890 (0.002) & 0.154 (0.001) & 0.895 (0.001)\\
      MGraphDTA~\cite{mgraphdta} & 0.233 (0.005) & 0.885 (0.004) & 0.150 (0.004) & 0.890 (0.002) \\
      iEdgeDTA~\cite{iedgedta} & 0.216 (0.004) & 0.897 (0.001) & 0.139 (0.001) & 0.890 (0.001)\\
      Mole-BERT~\cite{molebert} & 0.266 (-) & - & 0.157 (-) & -\\
      UniMAP~\cite{unimap} & 0.246 (-) & 0.888 (-) & 0.144 (-) & 0.891 (-) \\
      GDilatedDTA~\cite{gdilateddta} & 0.237 (-) & 0.885 (-) & 0.156 (-) & 0.876 (-)\\
      \midrule
      Pre-trained DTA & 0.205 (0.0008) & 0.893 (0.0021) & 0.162 (0.0012) & 0.866 (0.0004) \\
      \hdashline
      \quad +$k$NN-DTA & 0.190 (0.0004) & 0.905 (0.0021) & 0.146 (0.0004) & 0.886 (0.0004) \\
      \quad +Ada-$k$NN-DTA & 0.191 (0.0009) & 0.902 (0.0026) & 0.147 (0.0000) & 0.885 (0.0004) \\
      \bottomrule
  \end{tabular}
  % }
  \vspace{-0.2cm}
\end{table*}

\section{Experiments}
\label{sec:exp}
To evaluate our $k$NN-DTA, we first pre-train a neural DTA model as test model, and then perform the $k$NN retrieval. We introduce the experiments with different settings in this section. If not specified, the pre-trained model, the datastore creation, and the testset are all from the same domain. 

\subsection{Datasets and Pre-trained DTA Models}
We evaluate on four well-known DTA benchmarks, including BindingDB IC$_{50}$ and $K_i$~\citep{liu2007bindingdb}, DAVIS~\citep{davis2011comprehensive}, and KIBA~\citep{tang2014making}. Besides, there are four generalization testsets for zero-shot transfer learning. 
The statistics of these datasets are in the Appendix~\ref{append:data_detail}.

\noindent{\bf BindingDB}~\citep{liu2007bindingdb} is a database of different measured binding affinities.
Following previous works such as DeepAffinity~\citep{karimi2019deepaffinity} and MONN~\citep{li2020monn}, we evaluated on IC$_{50}$ and $K_i$ affinity scores with same data split, which are 60\% for training, 10\% for validation and 30\% for test. The label of affinity scores are transformed to logarithm scales as commonly done. 
To evaluate the zero-shot transfer ability of $k$NN-DTA, following \citep{karimi2019deepaffinity}, we test on four generalization testsets, ion channel/GPCR/tyrosine kinase/estrogen receptor, where the targets are not in the training set.
{\bf DAVIS}~\citep{davis2011comprehensive} contains selectivity assays of the kinase protein family and the relevant inhibitors with their respective dissociation constant ($K_d$) values. 
{\bf KIBA}~\citep{tang2014making} includes kinase inhibitor bioactivities measured in $K_i$, $K_d$, and IC$_{50}$, and the labels were constructed to optimize the consistency between them by using the statistics they embedded in these quantities~\citep{ozturk2018deepdta}. Following DeepPurpose~\citep{huang2020deeppurpose}, we split DAVIS and KIBA datasets into $7:1:2$ as train/valid/test sets.

To evaluate $k$NN-DTA, we first pre-train neural DTA models for each dataset, which consist of Transformer~\citep{vaswani2017attention} encoders and the upper Transformer layers for affinity prediction. The performance of these DTA models is ensured to be good when comparing to previous works. Then our $k$NN-DTA/Ada-$k$NN-DTA are performed upon the pre-trained DTA models. 
The details of the model architecture and training are put in the Appendix~\ref{append:model_config}.

\subsection{Parameters of $k$NN-DTA and Evaluation Metrics}
To find the best hyperparameters for $k$NN-DTA, we do search on each valid set. We tune $k, k_D, k_T$ in $[2^1, 2^2, ... 2^7]$, $\tau$, $\tau_D$ and $\tau_T$, in $[10^1,10^2,...,10^5]$, $\lambda$, $\lambda_D$ and $\lambda_T$ in $[0.1, 0.2,...,1.0]$. When searching neighbors, we use FAISS~\citep{johnson2019billion}, which is a library for efficient nearest neighbor search in high-dimensional spaces. The parameters for the best valid performance are applied to the test set. For training Ada-$k$NN-DTA, the hidden dimension of the meta-network is $32$ and we take no more than $5k$ steps training on one GPU on the valid data. 

We follow previous works~\citep{huang2020deeppurpose,ozturk2018deepdta,karimi2019deepaffinity} to evaluate the performance. Specifically, (a) root-mean-square error (RMSE) and (b) Pearson Correlation coefficient (R)~\citep{abbasi2020deepcda} are used to evaluate on BindingDB IC$_{50}$ and K$_i$ datasets, 
(c) mean-square error (MSE) and (d) Corcondance Index (CI)~\citep{gonen2005concordance} are on DAVIS and KIBA datasets.

\subsection{Results on BindingDB Benchmark}
The RMSE and Pearson Correlation results of BindingDB IC$_{50}$ and $K_i$ are shown in Table~\ref{tab:bindingdb_result}. For comparison, we take several works and existing best models as baselines,  including Random Forest~\citep{karimi2019deepaffinity}, DeepAffinity~\citep{karimi2019deepaffinity}, DeepDTA~\cite{ozturk2018deepdta}, MONN~\citep{li2020monn}, BACPI~\citep{li2022bacpi}, and SSM-DTA~\citep{pei2023breaking}. These baseline results are reported from original papers (Random Forest is reported in DeepAffinity and DeepDTA is reported in MONN).
From Table~\ref{tab:bindingdb_result}, we can see: 
(1) Comparing with existing works, our pre-trained DTA models achieve strong performances (e.g., $0.717$ RMSE), which already outperform the previous best BACPI~\citep{li2022bacpi} on both RMSE and R.
(2) After combining with our $k$NN-DTA, the performances can be further improved by a large margin. For instance, RMSE results on IC$_{50}$ and $K_i$ benchmarks are improved to $0.684$ and $0.750$, which significantly overpass the pre-trained models by $0.033$ and $0.035$ RMSE. 
(3) With Ada-$k$NN-DTA, the performances are further improved. The RMSE is reduced to $0.675$ and $0.735$.
Therefore, these numbers can clearly demonstrate the effectiveness of our $k$NN-DTA and also the adaptive learning of Ada-$k$NN-DTA. 

\subsection{Results on DAVIS and KIBA Benchmarks}
We then evaluate the performance on DAVIS and KIBA datasets, and the results are presented in Table~\ref{tab:davis_kiba_result}. Compared with BindingDB datasets, DAVIS and KIBA are relatively small-scale. The baseline methods are KronRLS~\citep{pahikkala2015toward}, GraphDTA~\citep{nguyen2021graphdta}, DeepDTA~\citep{ozturk2018deepdta}, DeepPurpose~\citep{huang2020deeppurpose}, DeepCDA~\cite{abbasi2020deepcda}, Affinity2Vec~\cite{thafar2022affinity2vec},
WGNN-DTA~\cite{jiang2022sequence}, and SSM-DTA~\citep{pei2023breaking}.
Again, we see that our pre-trained DTA models obtain good performances compared to previous best works, e.g. $0.205$ and $0.162$ MSE on DAVIS and KIBA respectively. By applying our $k$NN-DTA, MSE is reduced to $0.190$ and $0.146$. 
However, Ada-$k$NN-DTA performs similarly to the $k$NN-DTA. We then study the reason behind and find the shape of the probability density function for DAVIS/KIBA affinity is highly sharp and different from BindingDB (more details in the Appendix~\ref{append:data_detail}). 
We suspect this centralized distribution may hinder the learning effectiveness of the samples that are not specific and diverse enough. Nevertheless, Ada-$k$NN-DTA still achieves strong improvements upon the pre-trained DTA model. The results above also demonstrate the effectiveness of the $k$NN-DTA in enhancing DTA prediction.

\subsection{Retrieval from Other Datastore}
\label{sec:other_data}
\begin{table}[t]
% \begin{wraptable}{r}{7cm}
  \centering
  % \vspace{-0.5cm}
  \caption{Performance of the DAVIS pre-trained model with different datastores on DAVIS testset.}
  \label{tab:davis_out_of_domain_result}
  % \scalebox{0.9}{
  \begin{tabular}{lcc}
      \toprule
      \textbf{Method} & \textbf{MSE$\downarrow$} & \textbf{CI$\uparrow$} \\
      \midrule
      Pre-trained model & 0.207 & 0.896\\
      \hdashline
      \quad + $k$NN-DTA & 0.189 & 0.907\\
      \quad + $k$NN-DTA + BindingDB & 0.168 & 0.914 \\
      \quad + Ada-$k$NN-DTA + BindingDB & 0.168 & 0.916 \\
      \bottomrule
  \end{tabular}
  % }
\vspace{-0.5cm}
% \end{wraptable}
\end{table}

Apart from above experiments, we further verify whether adding other/external datastore for retrieval is beneficial. In this experiment, we take the pre-trained model on DAVIS. Besides the DAVIS training set as datastore, we also add BindingDB training data in the datastore, hence the datastore is from two different datasets. Note that part of the targets in the DAVIS are also in BindingDB, so this actually enlarge the retrieval datastore. The evaluation is performed on DAVIS testset and the results are presented in Table~\ref{tab:davis_out_of_domain_result}. We compare the $k$NN-DTA retrieval on DAVIS datastore, and DAVIS$+$BindingDB, and Ada-$k$NN-DTA on DAVIS$+$BindingDB. It can be seen that retrieval method benefits from additional data and improves the DTA performance, e.g., MSE is reduced from $0.189$ to $0.168$ when comparing the retrieval from DAVIS only with DAVIS+BindingDB. This experiment shows the easy adoption of our method and also the great potential in real applications. 

\section{Study}
To better understand our work, we conduct extensive studies. Without specific mention, we take BindingDB $K_i$ as testbed. 

\subsection{Ablation}

\begin{table}[t]
% \begin{wraptable}{r}{6.5cm}
  \centering
  % \vspace{-0.5cm}
  \caption{Ablation results (RMSE) on BindingDB $K_i$ dataset.}
  \label{tab:ablation}
  % \scalebox{0.9}{
  \begin{tabular}{lcc}
      \toprule
      \textbf{Method} & \textbf{Valid} & \textbf{Test} \\
      \midrule
      Pre-trained model & 0.795 & 0.784 \\
      % \midrule
      \hdashline
      $k$NN-DTA & 0.758 & 0.748 \\
      \quad - Label Aggregation  & 0.772 & 0.762 \\
      \quad - Representation Aggregation  & 0.763 & 0.753 \\
      \bottomrule
  \end{tabular}
  % }
  % \vspace{-0.3cm}
% \end{wraptable}
\end{table}

We first conduct ablation study to investigate the effect of our two aggregation ways. We remove the label aggregation and representation aggregation from our $k$NN-DTA separately and check the performance effect. In Table~\ref{tab:ablation}, we can see that (1) removing each of the two aggregation methods hurt the prediction performance. (2) Besides, both aggregation methods benefit the DTA prediction (each of the removed settings still outperforms pre-trained model). (3) Comparing these two methods, we can conclude that label aggregation contributes more to the success of $k$NN-DTA, e.g., the performance drop when removing label aggregation ($0.748$ v.s. $0.762$) is more than removing representation aggregation ($0.748$ v.s. $0.753$).

\subsection{Comparison with Biological Similarity}
\label{sec:chemical_similarity}
\begin{table}[t]
% \begin{wraptable}{r}{6cm}
  \centering
  % \vspace{-0.2cm}
  \caption{Retrieval methods comparison (RMSE) on BindingDB $K_i$ dataset with label aggregation.}
  \label{tab:chem_retrieval}
  \scalebox{1.0}{
  \begin{tabular}{lcc}
      \toprule
      \textbf{Method} & \textbf{Valid} & \textbf{Test} \\
      \midrule
      Pre-trained model & 0.795 & 0.784\\
      \midrule
      Chemical Retrieval & 0.776 & 0.764\\
      Embedding Retrieval & 0.763 & 0.753\\
      \bottomrule
  \end{tabular}
  }
  % \vspace{-0.4cm}
% \end{wraptable}
\end{table}

In drug discovery, a widely adopted way to retrieve similar molecule/protein is to use chemical/biological similarity measures, e.g., the 2D/3D structure similarity of the molecules/proteins. The most popular similarity measurement for molecules is Tanimoto similarity~\cite{fligner2002modification} based on the fingerprint. For protein, the common method is the normalized score of the Smith-Waterman (SW) alignment of the amino acid sequence~\cite{yamanishi2008prediction}, which compares segments of all possible lengths of the protein sequence and optimizes the similarity measure.
Hence, we first make a study to compare the retrieval cost by Tanimoto similarity and Smith-Waterman (SW) alignment score with our embedding similarity. For robustness, we use several drugs/targets as the queries and then count the average retrieval cost on the whole datastore (i.e. unique drugs/targets set of the whole set). We use the widely used toolkits, RDKit~\cite{landrum2013rdkit} and Biopython~\cite{cock2009biopython} to do molecule/protein similarity searches, and the fingerprint is RDKit topological fingerprint with 2048 bits.
To ensure a fair comparison, we use the CPU version for all packages, and the fingerprints of drugs are calculated and saved in advance (like our embedding datastore).
Averagely speaking, for each drug, the time costs for embedding and fingerprint similarity search are similar (millisecond order) as the fingerprint is vector indeed.
However, calculating the SW alignment score by Biopython for each target takes about hundreds of seconds, which is much slower compared with the protein embedding search (millisecond order).

Besides the cost comparison, we also care about the performance. Hence we make a further comparison of the embedding-based retrieval and the chemical/biological similarity retrieval. Due to the high cost of the chemical/biological similarity search with Biopython as shown in the above study, it is not practical to do a full set search. Thus, we reduce the searching space to $64$ using our embedding similarity search, then retrieve the nearest $32$ neighbors by these tools for statistic comparison. 
Then we evaluate $k$NN-DTA (for simplicity, we only use the label aggregation) by these two retrieval methods. The RMSE performance are shown in Table~\ref{tab:chem_retrieval}. We can see that our embedding-based retrieval shows a strong advantage over chemical/biological retrieval on prediction performance. 
The above comparisons of cost and performance demonstrate that our method is efficient and effective.

\begin{figure*}[t!]
     \centering
     \vspace{-0.5cm}
     \includegraphics[width=0.8\textwidth]{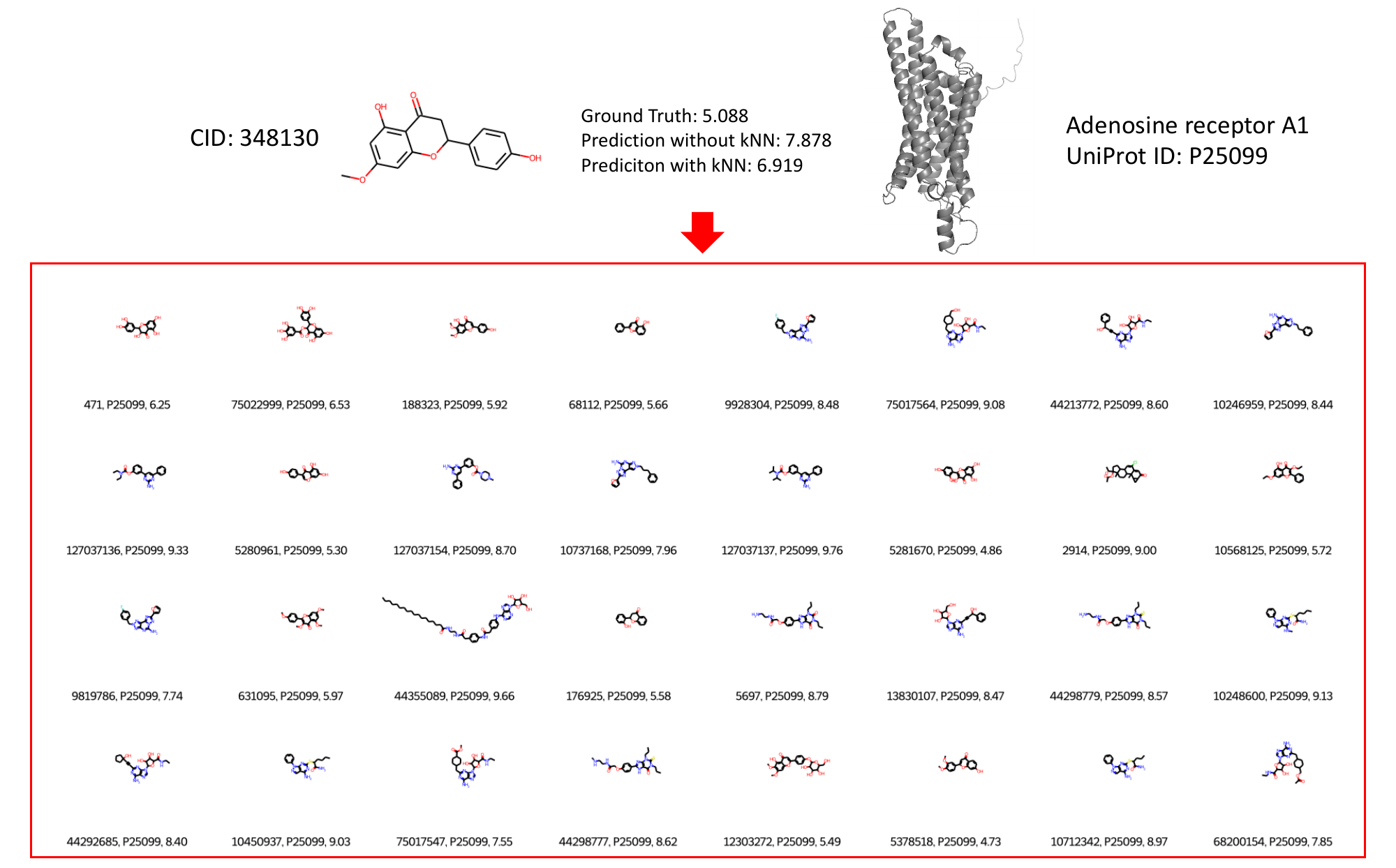}
         \caption{Case 1. Among these 32 neighbors, the target is the same for all neighbors. 
         }
     \label{fig:case}
     \vspace{-0.3cm}
\end{figure*}
\subsection{$k$NN-DTA for other Backbone models}
\begin{table}
% \begin{wraptable}{r}{6cm}
  \centering
  % \vspace{-0.75cm}
  \caption{Performance of $k$NN-DTA applied on different pre-trained DTA models on BindingDB $K_i$.}
  \label{tab:others}
  % \scalebox{0.82}{
  \begin{tabular}{lccc}
      \toprule
      \textbf{Pre-trained DTA} & \textbf{RMSE$\downarrow$} & \textbf{R$\uparrow$}\\
      \midrule
      12-layer pre-trained DTA & 0.854 & 0.844 \\
      \quad +$k$NN-DTA & 0.824 & 0.853 \\
      \midrule
      4-layer DTA train from scratch & 0.892 & 0.827 \\
      \quad +$k$NN-DTA & 0.872 & 0.836 \\
      \midrule
      BACPI~\cite{li2022bacpi} & 0.815 & 0.856 \\
      \quad +$k$NN-DTA & 0.797 & 0.863 \\
      \bottomrule
  \end{tabular}
  % }
  \vspace{-0.4cm}
% \end{wraptable}
\end{table}
Generally speaking, our $k$NN-DTA is model agnostic and it does not depend on specific architecture or what kind of pre-trained DTA model. Hence, in this subsection, we evaluate different pre-trained DTA models. Besides the Transformer network that is used as the DTA model in this paper, we also apply our $k$NN-DTA to graph neural network (GNN)-based DTA model prediction. 
Specifically, we first take the 12-layer pre-trained molecule/protein encoders and finetune them on DTA. We also take the 4-layer Transformer encoders that were trained from scratch for DTA prediction. Above two DTA models are still based on Transformer architecture but with different performances.
Besides, we take the recent best GNN work, BACPI~\cite{li2022bacpi}, as the DTA backbone model. Then we apply $k$NN retrieval on these different pre-trained DTA models to evaluate the performance on BindingDB $K_i$ test set. 
The results are shown in Table~\ref{tab:others}.
For the two Transformer-based DTA models, applying our $k$NN-DTA can consistently improve the model performance as we show in our main experiments.
For our reproduced BACPI, it achieves RMSE score $0.815$ and Pearson Correlation $0.856$, with $k$NN-DTA, the results are improved to $0.797$ RMSE and $0.863$ Pearson Correlation. 
These comparisons show the universal effectiveness of our $k$NN retrieval method. The method can improve performance not only on different model architectures but also on pre-trained DTA models with different performances.

\begin{table}[t]
  \centering
  \caption{Performance effect when varying the size of retrieval datastore on BindingDB $K_i$.}
  \label{tab:retrieval_size}
  \vspace{-0.3cm}
  \begin{tabular}{lccc}
      \toprule
      \textbf{Retrieval size} & \textbf{Valid} & \textbf{Test}\\
      \midrule
      Pre-trained model w/o $k$NN & 0.795 & 0.784 \\
      \midrule
      \quad +$k$NN-DTA on full set (1) & 0.758 & 0.748 \\
      \quad +$k$NN-DTA on half set (1/2) & 0.768 & 0.759 \\
      \quad +$k$NN-DTA on quarter set (1/4) & 0.770 & 0.760 \\
      \bottomrule
  \end{tabular}
\end{table}
\begin{table}[t]
  \centering
  % \vspace{-0.5cm}
  \caption{RMSE performance of the BindingDB $K_i$ pre-trained model on test sets with different visibility settings.}
  \label{tab:bindingdb_ki_seen_unseen}
  \vspace{-0.3cm}
  % \scalebox{0.9}{
  \begin{tabular}{lcccc}
      \toprule
      \multirow{2}{*}{\textbf{Setting}} & \textbf{Unseen} & \textbf{Unseen} & \textbf{Pre-trained} & \textbf{$k$NN}\\
      & \textbf{Drug} & \textbf{Target} & \textbf{Model} & \textbf{-DTA} \\
      \midrule
      Full Test & - & - & 0.784 & 0.748 \\
      \midrule
      \multirow{4}{*}{Visibility} & $\times$ & $\times$ & 0.744 & 0.702\\
      & $\checkmark$ & $\times$ & 0.813 & 0.782\\
      & $\times$ & $\checkmark$ & 1.443 & 1.397\\
      & $\checkmark$ & $\checkmark$ & 1.702 & 1.676\\
      \bottomrule
  \end{tabular}
\vspace{-0.2cm}
\end{table}

\subsection{Effect of The Size of Retrieval Datastore}
We further do another study to see the effect of the size of retrieval datastore. We conduct this study on BindingDB $K_i$ dataset, and we vary the datstore size from the full training data to half and quarter of the full set, then we evaluate the performance of the valid and test sets. The results are shown in Table~\ref{tab:retrieval_size}. From the table, we can clearly observe that the datastore size indeed impacts the final performance, but they all surpass the original model (without $k$NN retrieval). Generally, the larger the datastore is, the more possible that we can retrieve for similar drug-target pairs, and the larger performance improvement we can get.

\subsection{Results on Zero-shot Transfer}
\label{sec:zero-shot}
\begin{table*}[h]
  \centering
%   \advance\leftskip-0.5cm
  % \vspace{-0.1cm}
  \caption{RMSE/R performance evaluation of different methods on the BindingDB generalization testsets with IC$_{50}$ and $K_i$ metrics. `$x/y$': $x$ is the RMSE score and $y$ is the Pearson Correlation.}
  \label{tab:generalization_results}
%   \resizebox{1.1\linewidth}{!}{
  % \scalebox{0.8}{
  \begin{tabular}{l cc cc cc cc}
      \toprule
      \textbf{Dataset} & \multicolumn{2}{c}{\textbf{ER}} & \multicolumn{2}{c}{\textbf{Ion Channel}} & \multicolumn{2}{c}{\textbf{GPCR}} &  \multicolumn{2}{c}{\textbf{Tyrosin Kinase}} \\
      \midrule
      Method & IC$_{50}$ & $K_i$ & IC$_{50}$ & $K_i$ & IC$_{50}$ & $K_i$ & IC$_{50}$ & $K_i$ \\
      \midrule
      Random Forest & 1.41/0.26  & 1.48/0.14 & \textbf{1.24/0.16} & \textbf{1.46/0.21} & 1.40/0.25 & \textbf{1.20/0.19} & 1.58/0.11 & 1.75/0.10 \\
      DeepAffinity & 1.53/0.16 & 1.76/0.09 & 1.34/0.17 & 1.79/0.23 & 1.40/0.24 & 1.50/0.21 & 1.24/0.39 & 2.10/0.16  \\
      \midrule
      Pre-trained model & 1.42/0.38 & 1.40/0.29 & 1.47/0.13 & 1.50/0.27 & \textbf{1.39/0.31} & 1.31/0.38 & 1.26/0.48 & 1.54/0.40 \\
      \hdashline
      \quad +$k$NN-DTA & \textbf{1.41/0.40} & \textbf{1.34/0.36} & 1.46/0.13 & 1.49/0.27 & \textbf{1.39/0.31} & 1.30/0.38 & \textbf{1.22/0.49} & \textbf{1.51/0.40} \\
      \bottomrule
  \end{tabular}
  % }
  \vspace{-0.2cm}
\end{table*}
Experiments in Section~\ref{sec:exp} build datastores from the training set used for pre-trained model, and the testset is from same domain, which intuitively ensure the similarity between the datastore and the testset. To evaluate the generalization ability of our $k$NN-DTA, we conduct the following two additional experiments.

First, we conduct experiment on the BindingDB $K_i$ dataset, where the test set is segmented into four parts based on the visibility (seen or unseen) of drugs and targets during training. The RMSE results are shown in Table~\ref{tab:bindingdb_ki_seen_unseen}. Notably, $k$NN-DTA achieves consistent improvement over these four settings. This evidences kNN-DTA's potential in completely novel targets or drugs.

We further conduct a zero-shot transfer experiment on BindingDB generalization testsets. Specifically, the targets in ER/Ion Channel/GPCR/Tyrosin Kinase are hold out before data splitting, which are unseen and increasingly different from the BindingDB training set.
Thus, we take the model pre-trained on BindingDB, build the datastore on BindingDB training set, and then apply $k$NN-DTA to evaluate on these four testsets. The results of RMSE and Pearson Correlation are reported in Table~\ref{tab:generalization_results}. We can see that though these testsets are much different from the data in datastore, $k$NN-DTA also improves the performance on some specific sets. For instance, on Tyrosin Kinase IC$_{50}$ and $K_i$, the RMSE reduced $4$ and $3$ points.
Noting this zero-shot transfer is extremely hard. Thus, our method has potential towards the generalization ability. From these experiments, we can see this hard setting should be an important direction for future works.

\subsection{Embedding Visualization}
\label{sec:embed_visual}
\begin{figure}[t]
    \centering
    \includegraphics[width=0.5\textwidth]{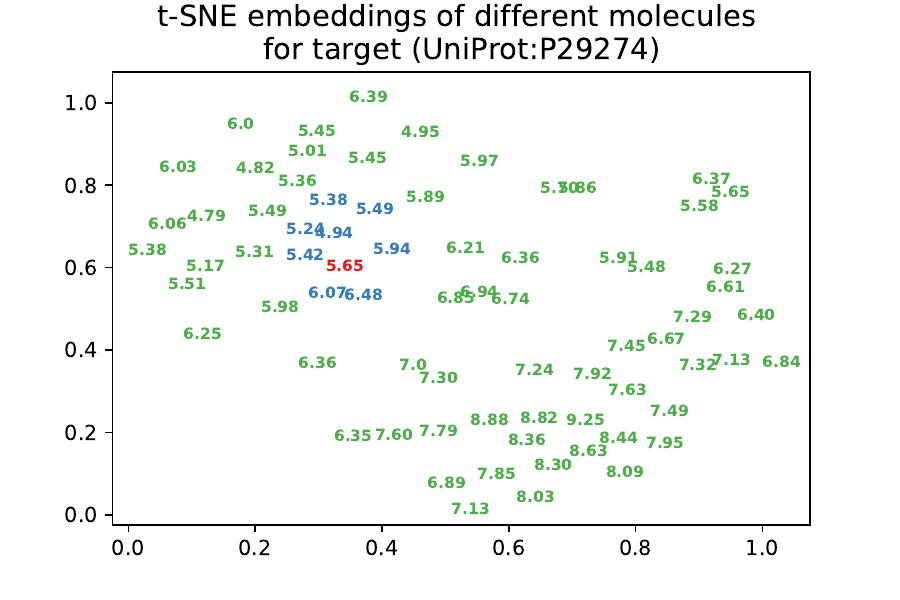}
    \vspace{-0.9cm}
    \caption{Embedding visualization for all the drugs that can bind to target (UniProt ID: P29274). The query drug (CID: 11791862) is in red, and the nearest $8$ drugs are in blue. The number of each node is the ground-truth affinity score.}
    \label{fig:embed_case}
    % \vspace{-0.6cm}
\end{figure}
Our retrieval-based method is mainly based on the assumption that for one drug-target pair, other drugs that are similar to the specific query drug may have similar affinity binding scores (e.g., point-wise drug retrieval). In order to better prove this assumption and demonstrate the effect of the retrieval-based method, we plot the embeddings for all drugs that can bind the P29274 (UniProt ID) target, the results are shown in Figure~\ref{fig:embed_case}.
The query drug (CID: 11791862) is in red color, and the nearest $8$ drugs are in blue color. The label for each node is its binding affinity score to the P29274 target. From the embedding visualization and the labeled affinity score, we can clearly observe that the nearest neighbors have similar affinity scores, especially when comparing to the right bottom drugs. Hence, this embedding visualization with affinity score can prove the assumption of our $k$NN retrieval method and support the motivation of our work.

\subsection{Case Study of Nearest Neighbors}
We finally provide some retrieved cases (more in the Appendix~\ref{append:case}) to better understand the effect of our method. The study is performed on pair-wise retrieval for simplicity. We randomly choose one sample that improves after applying our $k$NN-DTA. Then we look into their retrieved nearest pairs for study. 
We plot the paired cases with their drug (PubChem ID, graph visualization), target (UniProt ID, 3D visualization), and also their ground-truth binding affinity score ($K_i$), the pre-trained DTA predicted score and our $k$NN-DTA predicted score. For the retrieved neighbors of drug-target pairs ($k=32$), we show the graph visualization, PubChem ID of the drugs for clear understanding, and the UniProt ID of targets, also the affinity scores.
For the specific drug-target pair and their retrieved neighbors, we have several findings.
(1) For the retrieved neighbors, all of the pairs are with the same target, and the differences are from the drugs. This is reasonable since multiple drugs can be used for one target. We can also see from the visualized graphs that the retrieved drugs are in the similar structure, which further demonstrates that similar drugs can benefit our method for DTA prediction. (2) Our $k$NN-DTA model indeed helps the predicted affinity score to be closer to the ground-truth value, specifically for some out-of-distributed pairs.In Figure~\ref{fig:case}, the ground-truth values of the test samples are far different from the neighbors. The predictions from our pre-trained model are based on the training data so the predictions are also far from the ground-truth. With the help of neighbors by our $k$NN-DTA, the predicted values are pushed to be much closer to the ground-truth. This is interesting and demonstrates the value of $k$NN-DTA.

\section{Conclusions}
In this paper, we propose an embedding-based non-parametric retrieval method, $k$NN-DTA and its extension Ada-$k$NN-DTA, for drug-target binding affinity prediction so as to further exploit the potential upon an existing DTA model with no or light cost. 
Through a label aggregation with pair-wise embedding retrieval and a representation aggregation with point-wise embedding retrieval, $k$NN-DTA greatly benefits DTA prediction from these retrieved neighbors.
We verify the effectiveness of $k$NN-DTA on four benchmark sets (BindingDB IC$_{50}$ and $K_i$, DAVIS, KIBA), and obtain significant improvements over previous best models. 
Comprehensive studies and experiments prove the great potential/practicality of our work.
In the future, we will improve our method for better efficiency and also extend it to other applications for drug discovery.

\begin{acks}
This work was supported by the National Natural Science Foundation of China (NSFC Grant No. 62122089), Beijing Outstanding Young Scientist Program NO. BJJWZYJH012019100020098, and Intelligent Social Governance Platform, Major Innovation \& Planning Interdisciplinary Platform for the "Double-First Class" Initiative, Renmin University of China, the Fundamental Research Funds for the Central Universities, and the Research Funds of Renmin University of China.
Qizhi Pei is supported by the Outstanding Innovative Talents Cultivation Funded Programs 2023 of Renmin University of China.
\end{acks}

\bibliographystyle{ACM-Reference-Format}
\bibliography{knn_dta}

\appendix
\section{Experimental Settings}
\subsection{Dataset Details}
\label{append:data_detail}
\begin{figure*}
     \centering
     \begin{subfigure}[b]{0.45\textwidth}
         \centering
         \includegraphics[width=\textwidth]{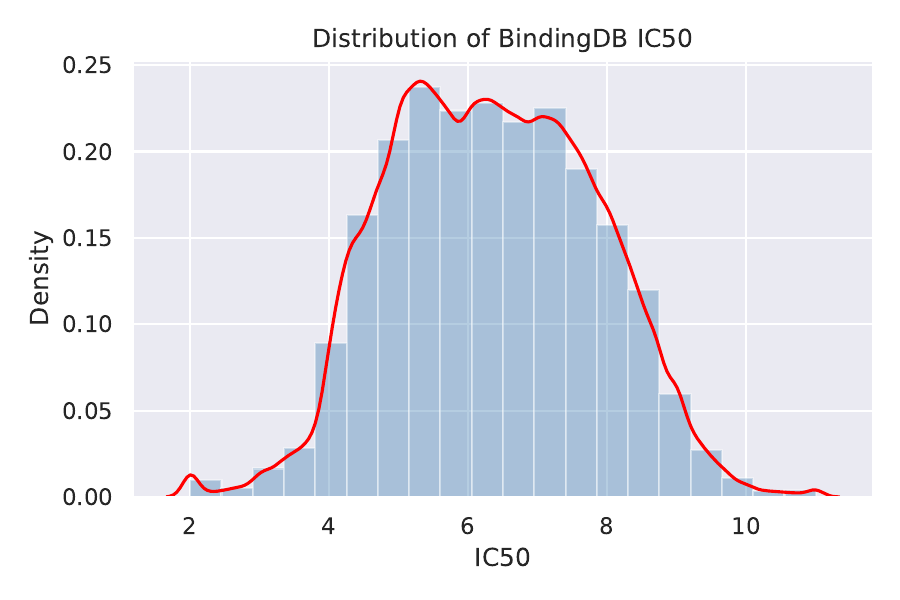}
         \caption{BindingDB IC$_{50}$ label distribution.}
         \label{fig:ic50_distribution}
     \end{subfigure}
     \begin{subfigure}[b]{0.45\textwidth}
         \centering
         \includegraphics[width=\textwidth]{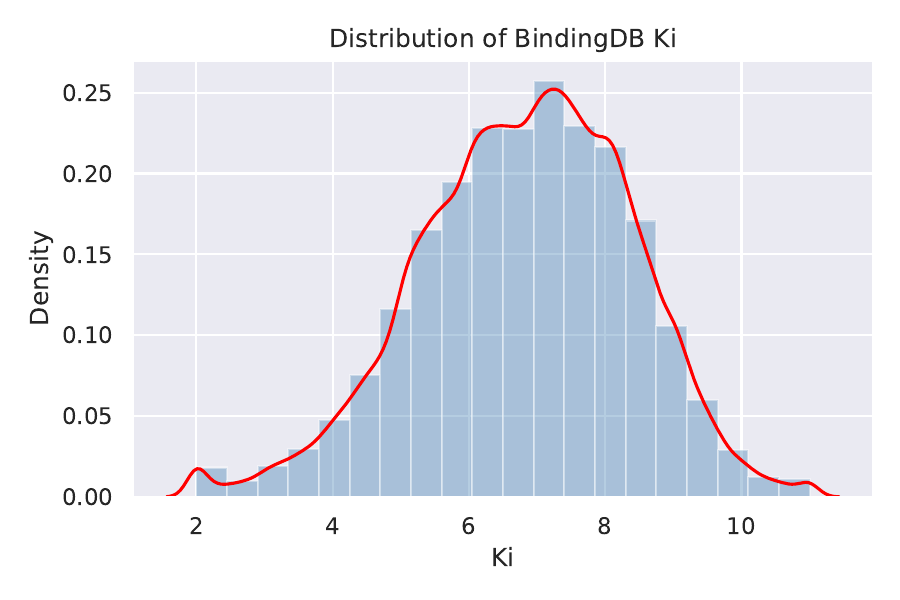}
         \caption{BindingDB $K_i$ label distribution.}
         \label{fig:ki_distribution}
     \end{subfigure}
     \begin{subfigure}[b]{0.45\textwidth}
         \centering
         \includegraphics[width=\textwidth]{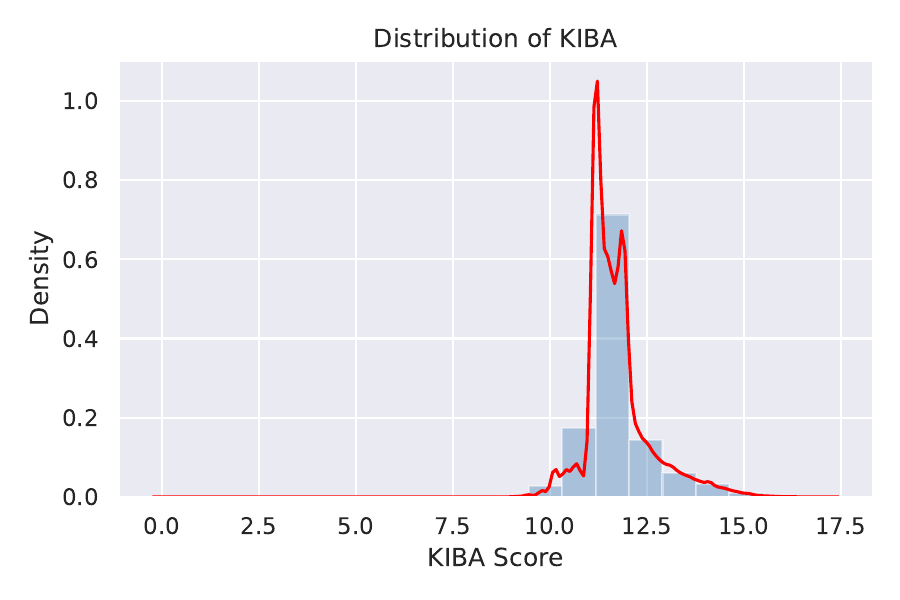}
         \caption{KIBA label distribution.}
         \label{fig:kiba_distribution}
     \end{subfigure}
     \begin{subfigure}[b]{0.45\textwidth}
         \centering
         \includegraphics[width=\textwidth]{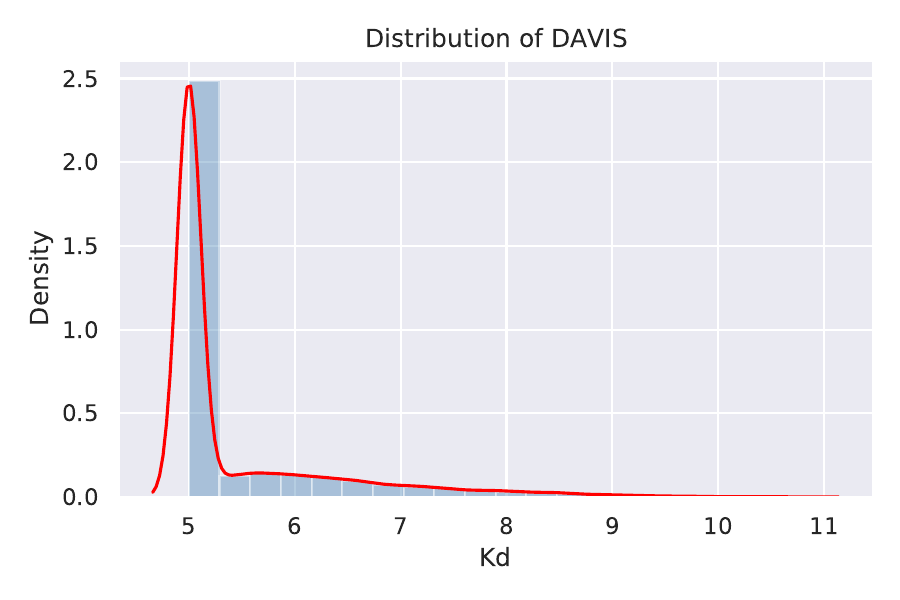}
         \caption{DAVIS label distribution.}
         \label{fig:davis_distribution}
     \end{subfigure}
     \caption{The label distribution of BindingDB IC$_{50}$ and $K_i$ datasets. The x axis is the affinity value (processed log version), and the y axis is the frequency ratio of the affinity value.}
      \label{fig:label_distribution}
\end{figure*}

The datasets we used for evaluation are BindingDB IC$_{50}$, $K_i$, KIBA, DAVIS, also the BindingDB generalization testsets. Besides, BindingDB $K_d$ dataset is used for out-of-domain datastore creation in Section 4.5. 
For BindingDB IC$_{50}$ and $K_i$, we randomly split them into train/valid/test with 6:1:3 as in~\citep{karimi2019deepaffinity}. For KIBA and DAVIS, train/valid/test sets are 7:1:2 as in~\citep{huang2020deeppurpose}. 
We give the detailed statistics of these datasets in Table~\ref{tab:dataset_detail} and BindingDB generalization testsets in Table~\ref{tab:bdb_gen}, including the number of drug-target pairs, the unique molecules and proteins. 
To better show the label information, we further give the label distribution plots of BindingDB IC$_{50}$, $K_i$, DAVIS and KIBA datasets in Figure~\ref{fig:label_distribution}. We can see that the affinity distributions of BindingDB are like normal distribution. However, the data distributions of DAVIS and KIBA are different, where the shape is sharp and the values are centered around specific area. This somehow hinders the learning ability of the Ada-$k$NN-DTA and affects the further performance gain of Ada-$k$NN-DTA on them.

\subsection{Model Configurations}
\label{append:model_config}
\begin{figure}[h]
    \centering
    \includegraphics[scale=0.5]{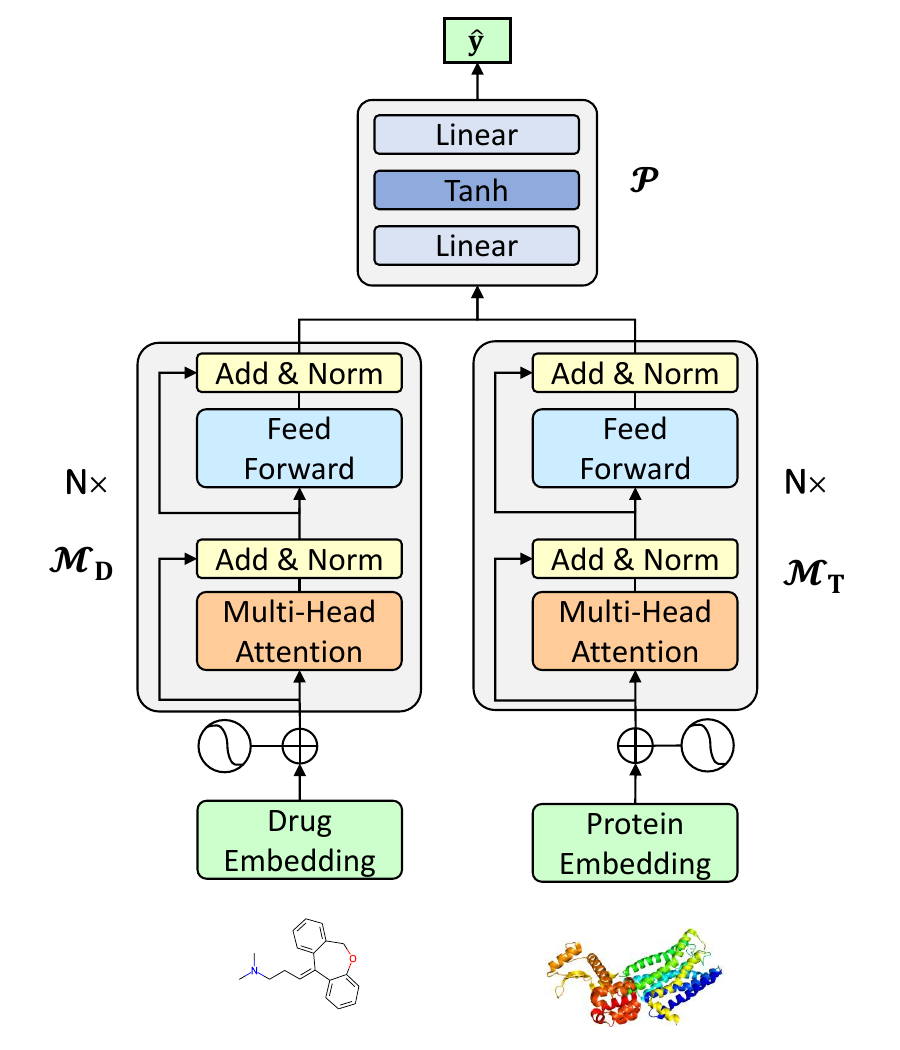}
    \caption{The architecture of our DTA prediction model, which contains one drug encoder and one target encoder ($\mathcal{M}_D$ and $\mathcal{M}_T$), and one upper prediction module ($\mathcal{P}$). Note that the first $12$ layers of $N=16$ layers encoder are pre-trained on unlabeled molecules and proteins and then fixed. Only the last $4$ layers are finetuned for DTA prediction.}
    \label{fig:model}
\end{figure}

To evaluate our method, we first pre-train DL-based DTA models for each dataset. 
We show the architecture of our DTA model in Figure~\ref{fig:model}.
We use two Transformer encoders for molecule encoder $\mathcal{M_D}$ and protein encoder $\mathcal{M_T}$ respectively, and each follows RoBERTa~\citep{liu2019roberta} architecture and configuration that consists of $16$ layers. The first $12$ layers of both encoders are initialized from the pre-trained molecule model and pre-trained protein model respectively. Specifically, the pre-trained molecule model is from a Transformer-based encoder that trained on molecules from PubChem~\citep{kim2021pubchem} dataset, and the pre-trained protein model is the same as the one in TAPE~\citep{rao2019evaluating} trained on proteins from Pfam~\citep{finn2014pfam} dataset (but we re-trained using Fairseq~\citep{ott2019fairseq}). As commonly done, both encoders take the masked language modeling objective for pre-training. The remained last $4$ Transformer layers are randomly initialized for $\mathcal{M_D}$ and $\mathcal{M_T}$. Then, the total $16$ layer encoders and an upper prediction module $\mathcal{P}$ are combined for DTA model training, which is the ``Pre-trained DTA" that we used for later $k$NN retrieval.
The embedding/hidden size and the dimension of the feed-forward layer are $768$ and $3,072$ respectively. The max lengths for molecule and protein are $512$ and $1,024$ respectively. 
The regression prediction head $\mathcal{P}$ is $2$-MLP layers with \texttt{tanh} activation function and the hidden dimension is $768$. 
During training, to save the computational cost, the first two pre-trained $12$-layer molecule and protein encoders are fixed and used as feature extractors, and only the last $4$ Transformer layers and $2$-MLP layers are learnable for DTA prediction. 
The implementation is based on Fairseq toolkit\footnote{\url{https://github.com/pytorch/fairseq}}.
The model is optimized by Adam~\citep{kingma2014adam} with a learning rate of $0.0001$. 
The dropout and attention dropout of two encoders are $0.1$. The learning rate is warmed up in the first $5\%$ update steps and then linearly decayed. The batch size is $32$ and we accumulated the gradients $8$ times during training. 

\begin{table}[htbp]
  \centering
  \caption{Dataset details.}
  \label{tab:dataset_detail}
%   \scalebox{0.9}{
  \begin{tabular}{lcccc}
      \toprule
      \textbf{Information} & \textbf{Pairs} & \textbf{Molecules} & \textbf{Proteins} \\
      \midrule
      BindingDB IC$_{50}$ & 376,751 & 255,328 & 2,782\\
      BindingDB $K_i$ & 144,525 & 87,461 & 1,620\\
      BindingDB $K_d$ & 7,900 & 63,233 & 1,504 \\
      KIBA & 118,254 & 2,068  & 229\\
      DAVIS & 30,056 & 68 & 379\\
      \bottomrule
  \end{tabular}
%   }
%   \vspace{-0.3cm}
\end{table}

\begin{table}[htbp]
  \centering
  \caption{BindingDB generalization testsets details.}
  \label{tab:bdb_gen}
%   \scalebox{0.9}{
    \resizebox{\linewidth}{!}{
  \begin{tabular}{lccccc}
      \toprule
      \multicolumn{2}{c}{\textbf{Dataset}}& \textbf{Pairs} & \textbf{Molecules} & \textbf{Proteins} \\
      \midrule
      \multirow{4}{*}{BindingDB IC$_{50}$} 
      & ER & 3,374 & 2,115 & 6\\
      & Ion Channel & 14,599 & 12,795 & 125\\
      & GPCR & 60,238 & 48,712 & 313\\
      & Tyrosin Kinase & 34,318 & 24,608 & 127\\
      \midrule 
      \multirow{4}{*}{BindingDB $K_i$} 
      & ER & 516 & 287 & 6\\
      & Ion Channel &  8,101 & 6,838 & 78\\
      & GPCR & 77,994 & 51,182 & 323\\
      & Tyrosin Kinase & 3,355 & 2,367 & 48\\
      \bottomrule
  \end{tabular}
  }
%   \vspace{-0.3cm}
\end{table}

\section{More Case studies}
\label{append:case}
\begin{figure*}[h]
     \centering
     \begin{subfigure}[b]{\textwidth}
         \centering
         \includegraphics[width=0.81\textwidth]{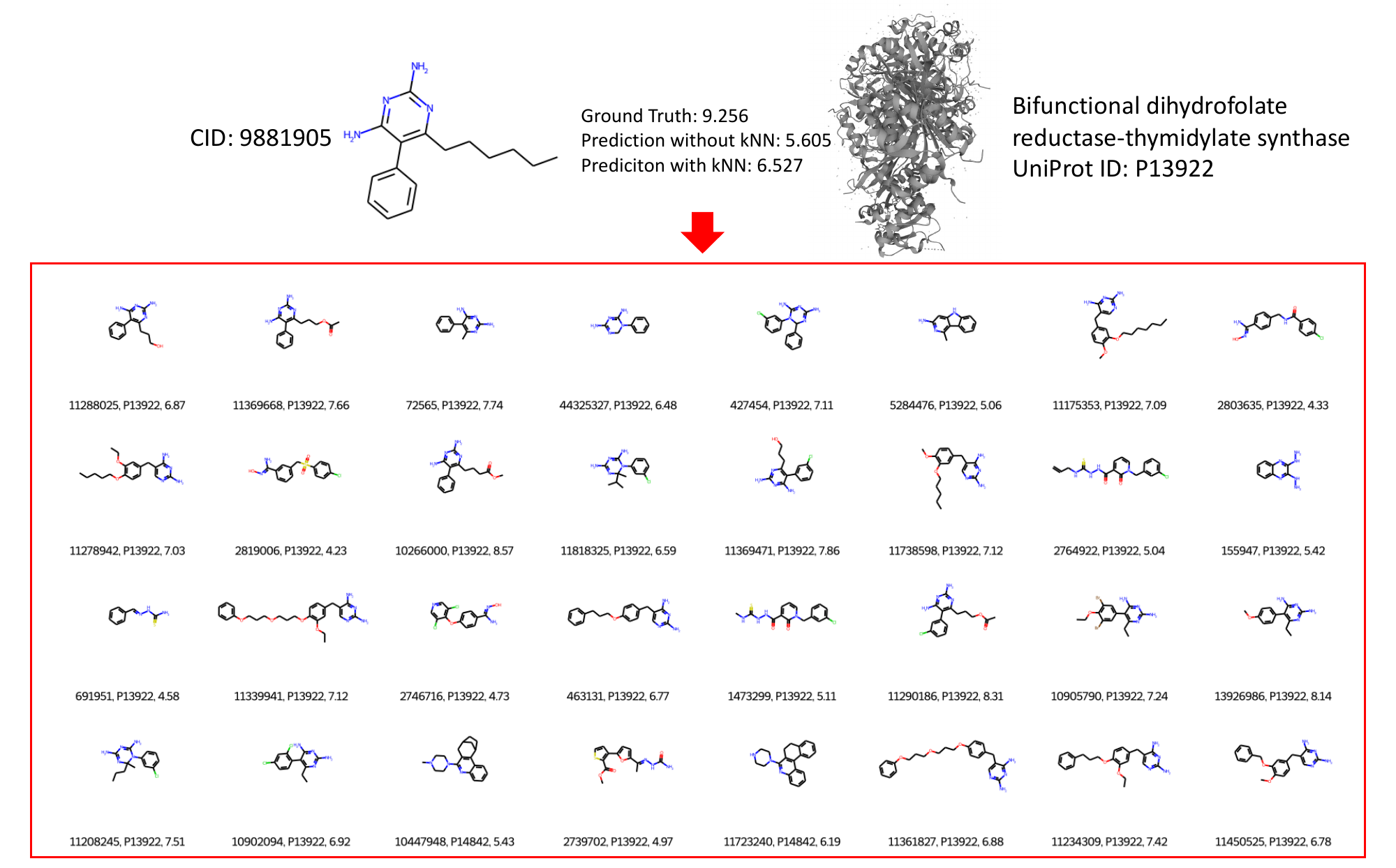}
         \caption{Case 2. Among these 32 neighbors, 30 are with the same target, and 2 are different.}
         \label{fig:case_2}
     \end{subfigure}
     \begin{subfigure}[b]{\textwidth}
         \centering
         \includegraphics[width=0.81\textwidth]{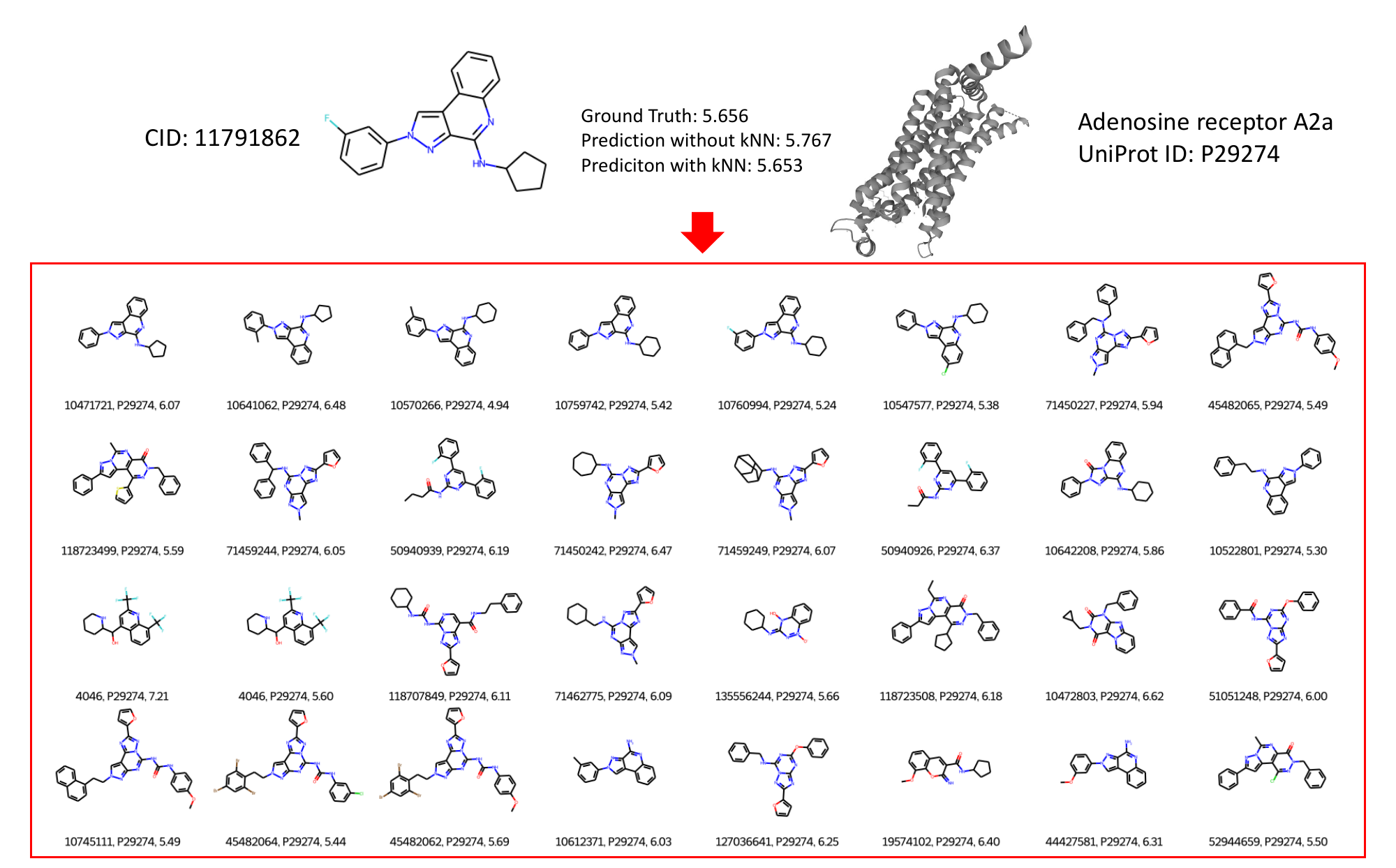}
         \caption{Case 3. Among these 32 neighbors, the target is the same for all neighbors.}
         \label{fig:case_3}
     \end{subfigure}
    %  \hfill
    \caption{Two cases of the test samples (top) and retrieved neighbors (bottom). The test sample includes the drug (PubChem ID, graph visualization), target (UniProt ID, 3D visualization), and their ground-truth binding affinity in $K_i$ measurement in log space, the pre-trained DTA predicted score and our $k$NN-DTA predicted score. The retrieved nearest neighbors include drug (PubChem ID, graph visualization), target (UniProt ID) and their binding affinity in $K_i$ measurement in log space.}
    \label{fig:three_cases}
\end{figure*}

We provide more cases about the retrieved nearest neighbors by the pair-wise retrieval method. We randomly choose some cases that benefit from our $k$NN-DTA method w.r.t the prediction performance. In Figure~\ref{fig:three_cases}, we plot the paired cases with their drug (PubChem ID, graph visualization), target (UniProt ID, 3D visualization), and also their ground-truth binding affinity score ($K_i$), the pre-trained DTA predicted score and our $k$NN-DTA predicted score. For the retrieved neighbors of drug-target pairs ($k=32$), we show the graph visualization, PubChem ID of the drugs for clear understanding, and the UniProt ID of targets, also the affinity scores. From these cases, we have several interesting findings. (1) For the retrieved neighbors, almost all of the pairs are with the same target, and the differences are from the drugs. This is reasonable since multiple drugs can be used for one target, and these pairs can help for the test sample. For instance, in case 1 (Figure~\ref{fig:case}) in the main paper, the target is adenosine receptor A1 and it shares with all retrieved neighbors. We can also see from the visualized graphs that the retrieved drugs are in the similar structure. (2) Our $k$NN-DTA model indeed helps the predicted affinity score to be closer to the ground-truth value, specifically for some out-of-distributed pairs. For example, we can see that in case 1 (Figure~\ref{fig:case}) in the main paper, the ground-truth values of the test samples are far different from the neighbors. The predictions from our pre-trained model are based on the training data so the predictions are also far from the ground-truth. With the help of neighbors by our $k$NN-DTA, the predicted values are pushed to be much closer to the ground-truth. This is interesting and demonstrates the value of $k$NN-DTA. For case 3 (Figure~\ref{fig:case_3}), though the prediction of pre-trained model is not far away from ground-truth (in-distribution), our $k$NN-DTA can make the prediction more accurate.

\end{document}